\documentclass[aps, prl, twocolumn, groupedaddress, 10pt, showpacs ,superscriptaddress]{revtex4-1}
\usepackage{pgfplots}
\usepackage{graphicx}
\usepackage{color}
\usepackage{amsfonts,amsmath,amsthm,amssymb,amscd}
\usepackage{dsfont}
\usepackage{mathtools}
\usepackage{amssymb}
\usepackage{braket}
\usepackage[colorlinks]{hyperref}
\hypersetup{
    citecolor = {blue},
    urlcolor = {purple}
}
\usepackage{tikz}
\usetikzlibrary{arrows, shadings, fadings, shadows, decorations.markings, shapes, positioning}

\def\ii{{\rm i}}
\newcommand{\dd}{{\rm d}}
\def\bra#1{\mathinner{\langle{#1}|}}
\def\ket#1{\mathinner{|{#1}\rangle}}

\definecolor{darkred}{rgb}{0.6,0.0,0}

\def\DD{{\mathcal{D}}}
\def\RR{{\mathcal{R}}}
\def\JJ{{\hat{\mathcal{J}}}}
\def\QQ{{\hat{\mathcal{Q}}}}
\def\qdef{{\mathfrak{q}}}

\newcommand{\be}{\begin{equation}}
\newcommand{\ee}{\end{equation}}
\newcommand{\ba}{\begin{aligned}}
\newcommand{\ea}{\end{aligned}}

\def\doi{http://dx.doi.org/}

\begin{document}

\title{Microscopic origin of ideal conductivity in integrable quantum models}

\author{Enej Ilievski}
\affiliation{Institute for Theoretical Physics Amsterdam and Delta Institute for Theoretical Physics,
University of Amsterdam, Science Park 904, 1098 XH Amsterdam, The Netherlands}

\author{Jacopo De Nardis}
\affiliation{D\'epartement de Physique, Ecole Normale Sup\'erieure,
PSL Research University, CNRS, 24 rue Lhomond, 75005 Paris, France}

\date{\today}

\begin{abstract}
Non-ergodic dynamical systems display anomalous transport properties. A prominent example are integrable quantum systems,
whose exceptional property are diverging DC conductivities.
In this Letter, we explain the microscopic origin of ideal conductivity by resorting to the thermodynamic particle content of a 
system. Using group-theoretic arguments we rigorously resolve the long-standing controversy regarding the nature of spin 
and charge Drude weights in the absence of chemical potentials.
In addition, by employing a hydrodynamic description, we devise an efficient computational method to calculate
exact Drude weights from the stationary currents generated in an inhomogeneous quench from bi-partitioned initial states.
We exemplify the method on the anisotropic Heisenberg model at finite temperatures for the entire range of anisotropies,
accessing regimes which are out of reach with other approaches.
Quite remarkably, spin Drude weight and asymptotic spin current rates reveal a completely discontinuous (fractal) dependence on the 
anisotropy parameter.
\end{abstract}


\maketitle


\paragraph*{Introduction.--}
Obtaining a complete and systematic understanding of how macroscopic laws of thermodynamics emerge from concrete microscopical models
has always been one of the greatest challenges of theoretical physics. Non-ergodic dynamical systems,
displaying a whole range of exceptional physical properties, have a special place in this context. One of their prominent features is
unconventional transport behaviour which attracted a great amount of interest after the authors of \cite{CZP95,ZNP97} conjectured that 
integrable quantum systems behave as ideal conductors. Although this has been shown to hold almost universally~\cite{ZNP97}, spin and 
charge transport in system with unbroken particle-hole symmetries instead show normal
(or even anomalous) diffusion~\cite{Sologubenko01,Hess01,Hlubek10,Maeter13,Hild14}.
Despite long efforts, the question whether the spin Drude weight in the isotropic Heisenberg spin chain
at finite temperature and at half filling is precisely zero is still vividly 
debated~\cite{CPC15,Karrasch17,CP17}, with a number of conflicting statements spread in the literature: while the prevailing opinion 
is that the spin Drude weight vanishes~\cite{Peres99,Zotos99,SPA11,Herbrych11,Znidaric11,Karrasch17,CP17}, other studies reach 
the opposite conclusion~\cite{Narozhny98,AG02,HM03,FK03,BFKS05,KBM12}.
As the question is inherently related to asymptotic timescales in thermodynamically large systems,
numerical approaches -- ranging from exact diagonalization to DMRG~\cite{AG02,HM03,Herbrych11,Znidaric11,SGB12,KBM12,Karrasch17} -- 
are insufficient to offer the conclusive and unambiguous answer.

In this Letter, we rigorously settle the issue by closely examining the underlying particle content which emerges
in thermodynamically large systems, and combine it with symmetry-based arguments to lay down the complete
microscopic background of ideal (dissipationless) conductivity.
Moreover, we present an efficient exact computational scheme for computing Drude weights with respect to general equilibrium states
by employing a nonequilibrium protocol based on hydrodynamic description developed in \cite{CDY16,BCNF16}.
Applying our method to the anisotropic Heisenberg model, we find that while the thermal Drude weight shows
continuous (smooth) dependence on anisotropy parameter, the spin Drude weight is a discontinuous function which
exhibit a striking fractal-like profile.\\

\paragraph*{Drude weights.--}
Transport behavior in the linear response regime is given by conductivity $\sigma^{(q)}(\omega)$ associated to charge density $q$.
The real part reads
\begin{equation}
{\rm Re}\,\sigma^{(q)}(\omega) = 2\pi\,\DD^{(q)}\delta(\omega) + \sigma_{\rm reg}^{(q)}(\omega),
\end{equation}
where $\sigma_{\rm reg}^{(q)}$ denotes the regular frequency-dependent part, whereas the magnitude of the singular part -- 
the so-called Drude weight $\DD^{(q)}$ -- signals dissipationless (ballistic) contribution.
The standard route to express $\DD^{(q)}$ is via Kubo formula, using the time-averaged
current autocorrelation function~\cite{Kubo57,Mahan_book}
\begin{equation}
\DD^{(q)}=\lim_{\tau \to \infty}\lim_{L\to \infty}\frac{\beta}{2\tau L}\int_{t=0}^{\tau}
\dd t\,\langle\JJ^{(q)}(t)\JJ^{(q)}(0)\rangle_{\beta,h},
\label{eqn:Green-Kubo}
\end{equation}
where $\langle \bullet \rangle_{\beta,h} = {\rm Tr}(\bullet\,\hat{\varrho}_{\beta,h})$,
$\hat{\varrho}_{\beta,h} \propto \exp{(-\beta \hat{H} + h\hat{N})}$,
denotes the grand canonical average at inverse temperature $\beta$ and chemical potential $h$~\footnote{
While the restriction to grand canonical ensembles $\hat{\varrho}_{\beta,h}$ adopted in this work is suitable
for studying $q=s,e$ (spin,energy) transport, our framework permits to consider completely general equilibrium states in a system.}
in a system of length $L$, while $\JJ^{(q)}=\sum_{i}\hat{j}^{(q)}_{i}$, where current densities $\hat{j}^{(q)}$ are determined
from local continuity equations, $\partial_{t} \hat{q}_{i} = \hat{j}^{(q)}_{i}-\hat{j}^{(q)}_{i+1}$.
While linear response formula \eqref{eqn:Green-Kubo} is suitable for efficient numerical simulations with DMRG
techniques~\cite{KBM13,KKM14,VKM16,Karrasch17}, it poses a formidable task for analytical approaches.
Spin Drude weight is commonly expressed via Kohn formula~\cite{Kohn64} (see also \cite{SS90,CZP95,ZNP97}) as 
the thermally averaged energy level curvatures under the application of a small twist $\phi$ (representing magnetic flux
piercing the ring),
$\DD^{(s)} = \frac{1}{2L}\sum_{n}w_{n}\partial^{2}_{\phi}E_{n}(\phi)\vert_{\phi=0}$,
with $w_{n}\propto \exp{(-\beta E_{n})}$ denoting the Boltzmann weights. Although Kohn formula proves convenient
for analytic considerations, it necessitates to properly resolve second-order system-size
corrections~\cite{FK98,Zotos99,Fujimoto99,BFKS05}. Alternatively, Drude weights may be
conveniently defined as the \emph{time-asymptotic rates} of the total current growth in the
zero-bias limit $\delta \mu_q \to 0$ (with $\mu_e = \beta$ and $\mu_s = h$, cf. Fig.~\ref{fig:profiles}),
\begin{equation}
\DD^{(q)} =  \lim_{\delta \mu_q\to 0}\lim_{t\to \infty}\lim_{L\to \infty}\frac{\beta}{2t}
\frac{  \langle \JJ^{(q)}(t;\delta \mu_q)\rangle_{\beta,h}}{\delta \mu_q}.
\label{eqn:D_noneq}
\end{equation}
This formulation was previously employed in \cite{VKM16} to study thermal transport in XXZ spin chain, and recently in
a DMRG study \cite{Karrasch17} of spin and thermal Drude weights in Hubbard and Heisenberg model. A related definition,
with the bias appearing as a Hamiltonian perturbation, was defined in \cite{IP12}, and shown to be equivalent
(under some mild assumptions) to Kubo formula \eqref{eqn:Green-Kubo}.

\begin{figure}[b]
\includegraphics[width=\hsize]{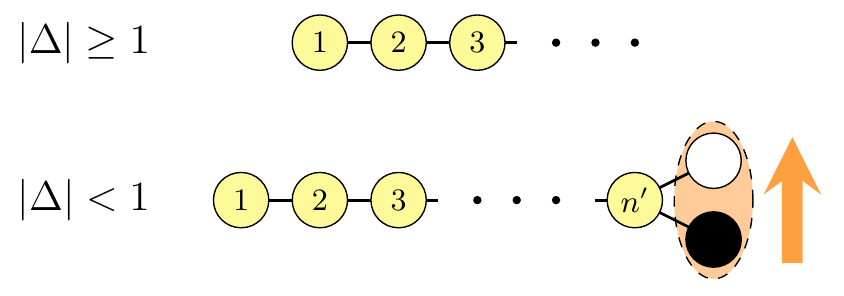}
\caption{Particle content of the XXZ Heisenberg model for $|\Delta|\geq 1$ (top) and $|\Delta|<1$ (bottom). While the former
consist of infinitely many bound magnons with densities $\rho_{a}(u)$, $a\in \mathbb{Z}_{\geq 1}$, the latter reduces to
$N_{\rm p}$ particles ($n^{\prime}=N_{\rm p}-2$) whose number depends discontinuously on $\Delta$.
Morphology of the graphs reflects how the particles effectively scatter among each other.
Black and white end nodes label a distinguished pair which forms a doublet with an
effective magnetic moment (orange arrow), being the \emph{only} particles in the spectrum which
transform non-trivially under the spin-reversal operation.}
\label{fig:particle_content}
\end{figure}
In ergodic dynamical systems, $\DD^{(q)} = 0$ is a consequence of the decay of dynamical correlations in Eq.~\eqref{eqn:Green-Kubo}. 
Integrable systems on the other hand feature stable \emph{interacting} particles,
representing collective thermodynamic excitations which undergo completely elastic (non-diffractive) scattering~\cite{ZZ79}, 
see Supplemental Material (SM) for further details~\cite{SM}. Such dynamical constraints result in a macroscopic number of conserved 
quantities $\QQ_{k}$ which prevent generic current-current correlations from completely decaying. This yields Mazur 
bounds~\cite{Mazur69,Suzuki71},
$\DD^{(q)} \geq \frac{1}{2L}\sum_{k}\langle \JJ^{(q)}\QQ_{k} \rangle^{2}_{\beta,h}/ \langle \QQ^{2}_{k}\rangle_{\beta,h}$,
which formally give exact results if \emph{all} extensive conserved quantities, satisfying
$\langle \QQ^{2}_{k} \rangle_{\beta,h} \sim \mathcal{O}(L)$, are included.
When $\JJ^{(q)}$ belongs to a conserved current, $[\JJ^{(q)},\hat{H}]=0$ (e.g. energy current $\JJ^{(e)}$ in the Heisenberg 
model~\cite{ZNP97,KS02}), the Drude weight is trivially finite and reads
$\DD^{(q)} = \lim_{L\to \infty}\tfrac{1}{2L}\langle (\JJ^{(q)})^{2}\rangle_{\beta,h}$.
Conversely, when $\JJ^{(q)}$ is not fully conserved, $\DD^{(q)}>0$ \emph{if and only if} there exist at least
one extensive conserved quantity $\QQ$ with a non-trivial overlap $\langle \JJ^{(q)}\QQ \rangle_{\beta,h} >0$. \\

\paragraph{Spin transport in the XXZ model.--}
We proceed by concentrating on the anisotropic Heisenberg model
\begin{equation}\label{eq:Hamiltonian}
\hat{H} = \sum_{i=1}^{L}\hat{S}^{x}_{i}\hat{S}^{x}_{i+1} + \hat{S}^{y}_{i}\hat{S}^{y}_{i+1} + \Delta 
(\hat{S}^{z}_{i}\hat{S}^{z}_{i+1}-\tfrac{1}{4}),
\end{equation}
in the entire range of anisotropy parameter $\Delta \in \mathbb{R}$.
For $|\Delta|>1$ ($|\Delta|\leq 1$) the thermodynamic spectrum is gapped (gapless).
We focus here on the elusive case of spin current $\JJ^{(s)}$.
The presence of chemical potential $h \neq 0$ which couples to $\hat{N}=\sum_{i}\hat{S}^{z}_{i}$ breaks particle-hole symmetry,
and renders $\DD^{(s)}>0$ for all $\Delta \in \mathbb{R}$ by virtue of a non-trivial Mazur bound~\cite{ZNP97}. At half filling
$h = 0$, however, the situation becomes more subtle. Since $\JJ^{(s)}$ is \emph{odd} under the spin-reversal transformation $\hat{R}=\prod_{i}\hat{S}^{x}_{i}$, 
namely $\hat{R}\,\JJ^{(s)}\,\hat{R} = -\JJ^{(s)}$, $\DD^{(s)}$ can only be finite if there exists an
extensive conserved quantity $\QQ$ of odd parity and finite overlap $\langle \JJ^{(s)}\QQ \rangle_{\beta,h} \neq 0$ \cite{ZNP97}.
In spite of substantial numerical evidence, clearly pointing towards $\DD^{(s)}>0$ for $|\Delta|<1$,
the long search for appropriate conservation laws only ended recently with a non-trivial bound obtained
in~\cite{Prosen11}, followed by a further improved bound derived in \cite{PI13} using a family of odd-parity 
charges stemming from non-compact representations of the quantized symmetry algebra $\mathcal{U}_{\qdef}(\mathfrak{sl}_{2})$.
Specifically, for commensurate values of anisotropy $\Delta=(\qdef + \qdef^{-1})/2$, where $\qdef = \exp{(\ii\pi m/\ell)}$
with $m<\ell$ ($\ell>2$) being two co-prime integers, the high-temperature bound (i.e. in the vicinity of $\beta \to 0$)
of \cite{PI13} reads explicitly 
\begin{equation}
\DD^{(s)} \geq \frac{\beta}{16} \frac{\sin^{2}{(\pi m/\ell)}}{\sin^{2}{(\pi/\ell)}}\left(1-\frac{\ell}{2\pi}\sin{(2\pi/\ell)}\right),
\label{eqn:analytic_bound}
\end{equation}
showing an unexpected `fractal' (nowhere-continuous) dependence on the anisotropy parameter $\Delta$.
At this stage, a few obvious
questions come to mind: (i) is the bound \eqref{eqn:analytic_bound} tight, or does it eventually smear out with the inclusion of extra 
(yet unknown) conservation laws? (ii) What is its value precisely at the isotropic point $\Delta = 1$ where the
bound \eqref{eqn:analytic_bound} becomes trivial? (iii) What is the physical origin  of the charges found
in \cite{Prosen11,PI13}. We subsequently provide natural and definite answers to these questions by expressing the Drude
weight in terms of balistically propagating particle excitations on the model.

\paragraph*{Particle content of the XXZ model.--}

Thermodynamic ensembles in integrable models are completely characterized by their particle content~\cite{YY69,Takahashi71,Gaudin71}.
Local statistical properties are encoded in \emph{macrostates}, corresponding to a complete set of mode density distributions
$\rho_{a}(u)$, where index $a$ labels distinct particle types, and $u$ is the rapidity variable which parametrizes particle momenta
$p_{a}(u)$. Distinct types of particles in the spectrum are intimately linked to representation theory of the underlying 
symmetry group. When $|\Delta|>1$, the thermodynamic spectrum of particles with respect to ferromagnetic vacuum consists of magnons 
($a=1$) and bound states thereof ($a\geq 2$)~\cite{Takahashi71,Gaudin71}. As explained in \cite{StringCharge}, these particle species 
are in a one-to-one correspondence with quantum transfer matrices composed of (auxiliary) finite-dimensional unitary irreducible 
representations of quantum group $\mathcal{U}_{\qdef}(\mathfrak{sl}_{2})$,
see \cite{SM}, and also~\cite{IMP15,IlievskiGGE15,IMPZ16_review,IQC17}.
Spin-reversal invariance of macrostates is a simple corollary of unitarity,
in turn implying that $\DD^{(s)}=0$ at $h=0$, in the entire range of anisotropies $|\Delta|\geq 1$.
We note that non-unitary highest-weight representations are of infinite dimension and do not enter into the description of
magnonic excitations.
In the critical regime $|\Delta|<1$ however, one finds an intricate situation where the particle content becomes unstable and
changes discontinuously upon varying $\Delta$~\cite{TS72}. When $\qdef$ is a root of unity, representing a dense set of points
in the interval $|\Delta| < 1$, the number of independent unitary transfer matrices and magnonic particles
both become finite. The latter represent $N_{\rm p}=\sum_{i=1}^{l}\nu_{i}$ bound excitations classified in~\cite{TS72} with aid of 
continued fraction representation,
$m/\ell = 1/(\nu_{1}+(1/\nu_{2}+\ldots))\equiv (\nu_{1},\nu_{2},\ldots,\nu_{l})$ (see SM for details~\cite{SM}),
which bijectively correspond to the finite-dimensional irreducible representations
of $\mathcal{U}_{\qdef}(\mathfrak{sl}_{2})$~\cite{StringCharge,DeLuca16}.
It is shown in \cite{DeLuca16} that the densities of a distinguished pair of
particles $\rho_{\bullet,\circ}$ (see Fig.~\ref{fig:particle_content}) map to the spectrum of the
odd-parity charges from \cite{PI13,Prosen14,Pereira14}, providing a link to
finite-dimensional \emph{non-unitary} representations of $\mathcal{U}_{\qdef}(\mathfrak{sl}_{2})$.
The lack of unitary implies that $\rho_{\bullet,\circ}(u)$ transform non-trivially under the spin-reversal transformation,
meaning that a change in the chemical potential $h$ only explicitly influences macrostates via the distributions 
$\rho_{\bullet,\circ}(u)$, while other densities get affected indirectly via interparticle interactions. The absence of exceptional 
particles in the $|\Delta|\geq 1$ regime on the other hand signifies that a macrostate is \emph{locally equivalent} to its
spin-reversed counterpart, and therefore no balistic spin transport between two regions with opposite magnetization density takes 
place.

\begin{figure}[t]
\includegraphics[width=\hsize]{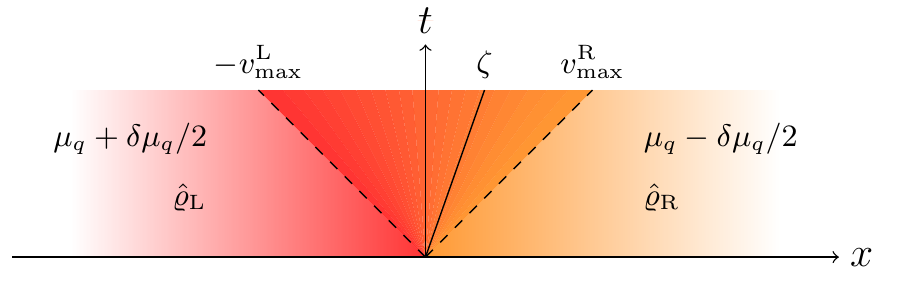}
\caption{Partitioning protocol: the initial state is prepared in two nearly identical grand canonical equilibria
$\hat{\varrho}_{\rm L,R}\propto \exp{(-\beta_{\rm L,R}\hat{H}+ h_{\rm L,R}\hat{S}^{z})}$,
representing a $q$-charged ($q=s,e$) domain wall of size $\delta q$, with the corresponding chemical potential drop
$\delta \mu_{q} = \mu_{q,\rm L}-\mu_{q,\rm R}$ (where $\mu_{e}= \beta$, $\mu_{s}= h$).
The initial defect expands in an inhomogeneous state localized within 
the `lightcone' $v^{\rm L}_{\rm max}<\zeta<v^{\rm R}_{\rm max}$. In the $t\to \infty$ limit, the state along each ray $\zeta=x/t$
relaxes in a quasi-stationary state which is uniquely characterized by particle distributions $\rho_{a}(u,\zeta)$,
for $a=1,\ldots,N_p$. Drude weight $\DD^{(q)}$ is proportional to the increment of the total current rate
$\lim_{t\to \infty}\mathcal{J}^{(q)}(t;\delta \mu_q)/t$ in the limit $\delta \mu_q\to 0$.}
\label{fig:profiles}
\end{figure}
\paragraph*{Drude weights from hydrodynamics.--}
We now describe a procedure for computing Drude weights using a nonequilibrium `partitioning protocol' developed
in \cite{CDY16,BCNF16}, drawing on the earlier ideas of \cite{Rubin71,Spohn77} and recent studies of 
CFTs~\cite{BD14,Bhaseen15,BD16_review}. A simple way to implement a thermodynamic gradient is to consider two 
partitions representing macroscopically distinct semi-infinite equilibrium states joined together at the point contact,
see Fig.~\ref{fig:profiles}. The imbalance at the junction induces particle flows between the two subsystems, with
a local quasi-stationary state emerging at late times along each ray $\zeta = x/t$. The latter is uniquely specified  
by the set of distributions $\rho_{a}(u,\zeta)$, pertaining to all types of particles in the spectrum
(labelled by $a=1,\ldots,N_p$), each obeying a local continuity equation \cite{CDY16,BCNF16}
\begin{equation}
\partial_t \rho_{a} (u,\zeta) + \partial_x \left[(v_{a}(u,\zeta) \rho_{a} (u,\zeta) \right]=0.
\label{eqn:equation_theta}
\end{equation}
Notice that, in distinction to non-interacting systems, particles' velocities $v_{a}(u)$ are dressed due to
interactions with a non-trivial background (macrostate)~\cite{YY69,Hubbard_book,FQ13},
$v_{a}(u) = \partial \omega_{a}(u)/\partial p_{a}(u)$,
where $\omega_{a}(u)$ and $p_{a}(u)$ are their dressed energy and momenta, respectively (see SM~\cite{SM}).
The solution of Eqs.~\eqref{eqn:equation_theta} for each ray $\zeta$  gives a family of densities $\rho_{a} (u,\zeta)$, 
see Fig.~\ref{fig:profiles}.

\begin{figure}[htb]
 \includegraphics[width=0.8\hsize]{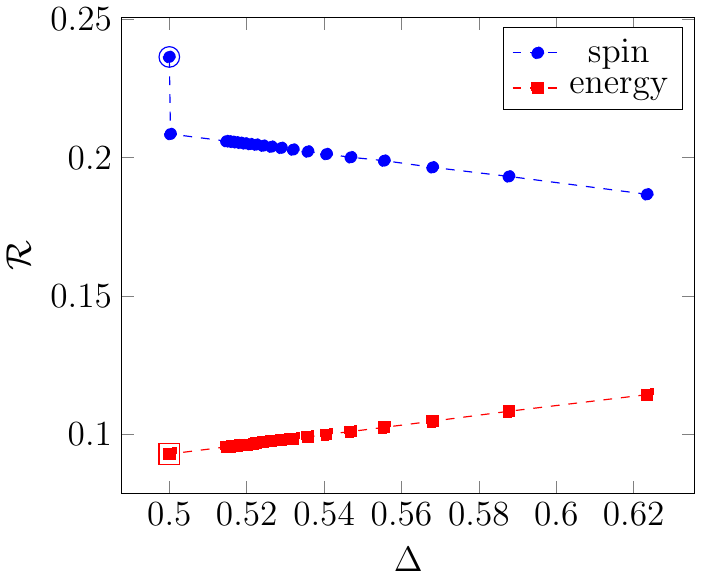}
\caption{Asymptotic spin (blue) and energy (red) current rates $\RR^{(q)} = \lim_{t\to \infty}\mathcal{J}^{(q)}(t)/ t =
\int_{-v^{\rm L}_{\rm max}}^{v^{\rm R}_{\rm max}}\dd \zeta\, j^{(q)}(\zeta;\delta \beta,\delta h)$,
emerging by joining two equilibrium states with chemical potentials $h_{\rm L,R} = \pm 1$ and inverse temperatures
$\beta_{L,R} = 1,3$. Considering the sequence $\Delta = \cos{(\pi/(3 + 1/\nu_{2}))}$ for $\nu_2 =2, 3, \ldots, 20,10^{3}$
(the points at $\nu_2=10^{3}$ are obtained by linearly extrapolation of other $\nu_{2}$-points), we demonstrate
that $\lim_{\nu_{2}\to \infty}\DD^{(s)}\neq \DD^{(s)}(\gamma=\tfrac{\pi}{3})$ (open circle). The same holds in general
when approaching a value of $\gamma$ parametrized by $l-1$ integers $\nu_{i}$ as the $\nu_{l}\to \infty$ limit of the
order-$l$ sequence of $\nu_{i}$. This indicates that spin current is a nowhere-continuous function of $\Delta$ within $|\Delta|<1$
(cf. Eq.~\eqref{eqn:analytic_bound} for the exact analytic high-temperature results).
Unlike $\RR^{(s)}$, the thermal current rate $\RR^{(e)}$ depends continuously on $\Delta$, as shown for $\nu_{1}=3$ (open square).}
\label{fig:fractality_non_eq}
\end{figure}

\begin{figure}[b]
\includegraphics[width=0.85\hsize]{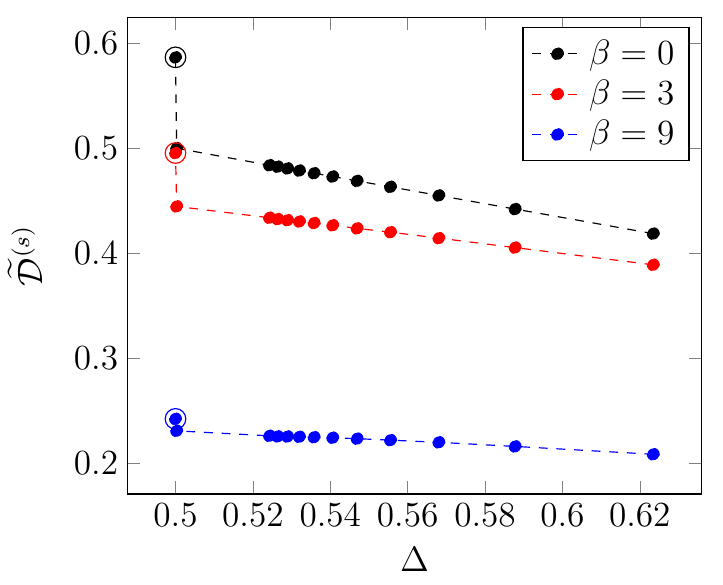}
\caption{Rescaled spin Drude weight $\widetilde{\DD}^{(s)} = (16/\beta)\DD^{(s)}$ obtained from
Eq.~\eqref{eqn:D_hydro} for various temperatures at $\Delta = \cos{(\pi/(3 + 1/\nu_2))}$, for
$\nu_2 = \{2, 3, \ldots, 12,10^3\}$ ($\nu_2=10^3$ are obtained by linear extrapolation of other $\nu_{2}$-points),
and for $\Delta = \cos(\pi/3 )=0.5$ (open circles). For a dense set of commensurate anisotropies
$\Delta = \cos{(\pi m/\ell)}$, $\DD^{(s)}$ is found to be a discontinuous function of $\Delta$ at arbitrary finite temperature
(see explanation in the caption of Fig.~\ref{fig:fractality_non_eq}).}
\label{fig:fractality}
\end{figure}

Computing the Drude weights requires infinitesimal gradients.
We thus consider two thermodynamic subsystems prepared in almost identical equilibrium states which differ
by a slight amount $\delta q$ in the charge density $q$ and experience a chemical potential jump $\delta \mu_q$ at the contact.
Transforming \eqref{eqn:D_noneq} to the lightcone frame, we find
\begin{equation}\label{eqn:D_hydro}
\DD^{(q)} = \lim_{\delta \mu_q \to 0}\frac{\beta}{2\,\delta \mu_q }
\int_{-v^{\rm L}_{\rm max}}^{v^{\rm R}_{\rm max}}\dd \zeta\, j^{(q)}(\zeta;\delta \mu_q),
\end{equation}
where $j^{(q)}(\zeta;\delta \mu_q)$ designates the quasi-stationary expectation value of the current density
in the direction of $\zeta$ emanating from the contact.
Using the hydrodynamical approach, we first verified the infinite-temperature results of Eq.~\eqref{eqn:analytic_bound}, and found 
perfect numerical agreement
(with absolute precision $<10^{-4}$), at $\Delta = \cos{(\pi m/\ell)}$ for different values of $\nu_1,\nu_2,\nu_3$.
We subsequently confirmed the discontinuous nature of the spin Drude weight as function of $\Delta$ not only at
infinite temperature~\cite{PI13}, but also at finite temperatures $\beta^{-1}$. As temperature is lowered, the discontinuities of 
$\DD^{(s)}$ become less pronounced (see Fig.~\ref{fig:fractality}), while as $T\to 0$ we find a power-law behavior
$\DD^{(s)} - \DD^{(s)}(T=0) \sim T^{2/(\ell/m-1)}$ (see Fig. 3 in~\cite{SM}).
Moreover, the hydrodynamic description remains applicable at finite bias $\delta \mu_q$, hence allowing to probe quantum 
transport properties even in the non-linear regime~\cite{BF16,CDY16,BCNF16,DeLuca16} and revealing that the asymptotic
spin-current rate $\RR^{(s)}=\lim_{t\to \infty}\mathcal{J}^{(s)}(t)/t$ is (unlike e.g. energy current rate $\RR^{(e)}$) an everywhere
discontinuous function of anisotropy $\Delta$, see Fig.~\ref{fig:fractality_non_eq}. Additional figures, showing low-temperature
behaviour of $\DD^{(s)}$ and its dependence on chemical potential $h$, are given in SM~\cite{SM}.
\\

\paragraph{Hubbard model.--}

A situation analogous to that of the isotropic Heisenberg model occurs in the (fermionic) Hubbard 
model~\cite{Takahashi72,Hubbard_book}, where in spite of solid evidence in favor of the vanishing finite-temperature
spin and charge Drude weights $\DD^{(c,s)}=0$ in the absence of the respective chemical potentials
(see \cite{Peres99,KEH99,Carmelo13,KKM14}), the definite conclusion is still lacking~\cite{Karrasch17,KPH16}.
A possibility of having additional (unknown) odd-parity conservation laws can however now be quickly ruled out by
invoking group-theoretic arguments along the same lines of the isotropic Heinsenberg model. In Hubbard model,
the entire space macrostates is in a one-to-one correspondence with particle-hole invariant commuting (fused) transfer matrices,
pertaining to a discrete family of \emph{unitary} irreducible representations of the underlying quantum symmetry~\cite{Cavaglia15}.
This readily implies vanishing finite-temperature charge/spin Drude weights $\DD^{(c,s)}=0$ when the corresponding chemical potentials
vanish, irrespective of the interaction strength. In the presence of external potentials the Drude weights are known to take 
finite values by virtue of Mazur bounds, cf.~\cite{ZNP97}. As the particle content of Hubbard model is robust against varying the 
coupling strength, the Drude weights exhibit a continuous dependence on it.

\paragraph*{Conclusions.--}
We presented a rigorous and intuitive picture for understanding the phenomenon of ideal conductivity in generic
integrable quantum models. Dissipationless transport of generic local charges is shown to be directly linked to the
interacting particles of a theory.
Nonetheless, spin (or charge) Drude weights in particle-hole symmetric models in the half-filled regimes
show exceptional behavior and require a careful analysis by examining the particle content of the model.

While our framework is applicable in general, we focused on the interesting case of the
anisotropic Heisenberg model. In the gapped phase, $|\Delta|\geq 1$, particles correspond to an infinite 
hierarchy of magnonic bound states which are robust under varying the anisotropy parameter~\cite{Takahashi71,Gaudin71}.
The fact that the corresponding particle density operators are insensitive to flipping the spins implies that two thermodynamic states
which are characterized terms of mode occupation distributions are (locally) identical, and no ballistic flow of particles across
the magnetic domain wall at zero magnetization density can occur.
Within the interval $|\Delta|<1$ however, the particle content for commensurate values of $\Delta$ consists of finitely many 
particles whose number depends discontinuously on $\Delta$~\cite{TS72}.
In this case, ballistic \emph{spin} transport is enabled by the appearance of a distinguished pair of particles which are
not invariant under the spin reversal and hence allow for chiral (i.e. spin-carrying) states. It should be stressed that the above
qualitative picture can be established independently from any quantitative analysis.

By employing a nonequilibrium partitioning protocol, we presented an exact numerical computation of Drude weights and
applied it on the anisotropic Heisenberg spin chain.
Our results rigorously prove that the formal infinite-temperature bound derived in \cite{PI13} is the exact Drude weight at infinite 
temperature, and moreover that the time-asymptotic spin current rate in the XXZ chain is a nowhere-continuous function of $\Delta$ for any finite temperature and even in the non-linear regime.
These observations indicate that the physics in the gapless regime $|\Delta|<1$ depends abruptly on the `commensurability effect', 
resembling the  pattern found in famous Hofstadter butterfly~\cite{Hofstadter76,WZ94,Faddeev95} multifractal spectrum.
As a future task, it would be valuable to perform high-precision finite-time numerical analysis to determine whether the `fractality'
can be detected via anomalously large relaxation times.

A number of intriguing open problems remain. Most notably, understanding the microscopic mechanism underlying
normal or anomalous diffusion which typically coexists with the ballistic channel,
see e.g. \cite{Fabricius98,SPA09,Langer09,Znidaric11,Jesenko11}), and recent work \cite{MKP17,LZP17}.
Another open question is to explain diffusive behavior in the semi-classical regime of the Heisenberg ferromagent~\cite{PZ13}, 
governed by Landau--Lifshitz action~\cite{FT_book} whose solitons are identified as long-wavelength macroscopic
bound states~\cite{Sutherland95}.

\paragraph*{Note added.} After this work appeared online, an independent work \cite{BVKM17} which partially overlaps with this
Letter also shows that the spin Drude weight could be obtained from hydrodynamics.

\paragraph*{Acknowledgements.}
The authors are grateful to E.~Quinn for fruitful discussions and comments on the manuscript, and T.~Prosen for valuable feedback.
E.I. acknowledges support by VENI grant number 680-47-454 by the Netherlands Organisation for Scientific Research (NWO).
J.D.N. acknowledges support by LabEx ENS-ICFP:ANR-10-LABX-0010/ANR-10-IDEX-0001-02 PSL*.

\bibliography{Drude}

\pagebreak

\appendix
\onecolumngrid

\clearpage

\begin{center}
{\large{\bf Supplemental Material for\\
``Microscopic origin of ideal conductivity in integrable quantum models"}}
\end{center}

This Supplemental Material contains a short review of the particle content in the anisotropic Heisenberg model,
along with the description of their dressed energies and momenta, and explicit construction of the corresponding
density operators. A few additional figures, showing temperature and chemical potential dependence of the spin
Drude weight, are provided as well.

\section{Particle content of the anisotropic Heisenberg model}
The anisotropic (XXZ) Heisenberg Hamiltonian,
\begin{equation}
\hat{H} = \sum_{i=1}^{L}\hat{S}^{x}_{i}\hat{S}^{x}_{i+1} + \hat{S}^{y}_{i}\hat{S}^{y}_{i+1} +
\Delta (\hat{S}^{z}_{i}\hat{S}^{z}_{i+1}-\tfrac{1}{4}),
\end{equation}
is diagonalized by Bethe Ansatz. We first assume $|\Delta|>1$ and parametrizing $\Delta = \cosh{(\eta)}$.
Any eigenstate in a finite system of size $L$ is assigned a unique set of rapidities $\{u_{k}\}^{M}_{k=1}$ (equiv. a set
of quantum numbers), defined from solutions of Bethe equations
\begin{equation}
e^{\ii p(u)L}\prod_{k=1}^{M}S_{11}(u-u_{k}) = -1\qquad {\rm for}\quad u\in \{u_{k}\}^{M}_{k=1},
\label{eqn:BE}
\end{equation}
where $M$ is the number of down-turned spins which relates to magnetization $S^{z}=L/2-M$.
The (complex) solutions given by $\{u_{k}\}^{M}_{k=1}$ referred to as the Bethe roots.
In the thermodynamic limit, defined by taking $L\to \infty$ and $M\to \infty$ limits (keeping ratio $M/L$ finite),
the solutions to Bethe equations organize into regular patters which indicate the presence of well-defined particle excitations.
Presently, these correspond to magnons and their bound states, known also in the literature
as the Bethe strings~\cite{Takahashi71}. A `$k$-string' solution reads
\begin{equation}
\{u^{k,m}_{\alpha}\} = \left\{u^{k}_{\alpha} + (k+1-2m)\tfrac{\ii \eta}{2}\right\},\qquad m=1,2,\ldots,k,
\end{equation}
where label $\alpha$ enumerates distinct $k$-strings and $m$ runs over their internal rapidities.
Scattering amplitudes associated to magnonic particles are
\begin{equation}
S_{j}(u) = \frac{\sin{(u-j\tfrac{\ii \eta}{2})}}{\sin{(u+j\tfrac{\ii \eta}{2})}},\qquad
S_{jk}(u) = \prod_{m=-\tfrac{k-1}{2}}^{\tfrac{k-1}{2}}\prod_{n=-\tfrac{j-1}{2}}^{\tfrac{j-1}{2}}S_{2m+2n+2} =
S_{|j-k|}S_{j+k}\prod_{m=1}^{{\rm min}(j,k)-1}S^{2}_{|j-k|+2m},
\label{eqn:scattering_amplitudes}
\end{equation}
using convention $S_{0}\equiv 0$. In the $L\to \infty$ limit, particle rapidities become densely distributed along the
real axis in the rapidity plane. This permits to introduce distributions $\rho_{k}(u)$ of $k$-string particles, along with
the dual hole distributions $\bar{\rho}_{k}(u)$ (holes are by definition solutions to Eq.~\eqref{eqn:BE} which differ from
Bethe roots $u_{k}$). The quantization condition \eqref{eqn:BE} gets accordingly replaced with the integral
Bethe--Yang equations~\cite{YY69}
\begin{equation}
\rho_{j} + \bar{\rho}_{j} = a_{j} - a_{jk}\star \rho_{k},
\end{equation}
where the kernels are given by the derivatives of the scattering phases,
\begin{equation}
a_{j}(u) = \frac{1}{2\pi \ii}\partial_{u}\log S_{j}(u),\qquad
a_{jk}(u) = \frac{1}{2\pi \ii}\partial_{u}\log S_{jk}(u).
\label{eqn:scattering_kernels}
\end{equation}
Here and below we use a compact notation for the convolution integral, $(f\star g)(u)=\int_{\mathbb{R}}f(u-t)g(t)\dd v$,
and repeated indices are summed over. On each rapidity interval $[u,u+\dd u]$ there is a macroscopic number of microstates,
for which a combinatorial entropy density per mode reads~\cite{YY69}
\begin{equation}
s_{j}(u) = \rho_{j}\log\left(1+\frac{\bar{\rho}_{j}(u)}{\rho_{j}(u)}\right)+
\bar{\rho}_{j}\log\left(1+\frac{\rho_{j}(u)}{\bar{\rho}_{j}(u)}\right).
\end{equation}

\paragraph{Isotropic point.}
The above results are extended to the isotropic point $\Delta=1$ after taking a scaling limit $u\to u\eta$, and subsequently
sending $\eta\to 0$.

\subsection{Gapless regime}
Classification of particle types in the gapless regime $|\Delta|<1$ can be found in \cite{TS72}. Here, in addition to the magnon type 
label $k$, an extra parity label $v\in \pm$ is required. Importantly, integers $k$ now no longer coincide with the length of a
string, i.e. a number of magnons forming a bound state.
Instead, the $k$-th particle consists of $n_{k}$ Bethe roots and carries parity $v_{k}$
(see \cite{TS72} for further details). Setting $\Delta = \cos{(\gamma)}$, where $\gamma/\pi = m/\ell$ (with $m,\ell$ co-prime
integers) is a \emph{root of unity}, the number of distinct particles in the spectrum is \emph{finite}. Changing the parametrization
$u\to \ii u$, $\eta\to \ii \gamma$ and incorporating the additional parity label, the elementary scattering amplitudes and
kernels read
\begin{equation}
S_{k}(u)\to S_{(n_{j},v_{j})}(u) = \frac{\sinh{(u - n_{j}\tfrac{\ii \gamma}{2}+(1-v_{j})\tfrac{\ii \pi}{4})}}
{\sinh{(u + n_{j}\tfrac{\ii \gamma}{2}+(1-v_{j})\tfrac{\ii \pi}{4})}},
\end{equation}
whereas the full set of scattering kernels are (likewise for the $|\Delta|>1$ case) obtained by fusion
(cf. Eqs.~\eqref{eqn:scattering_amplitudes} and \eqref{eqn:scattering_kernels}).
The Bethe--Yang equations for string get slightly modified, reading
\begin{equation}
\sigma_{j}(\rho_{j}+\bar{\rho}_{j}) = a_{j} - a_{jk}\star \rho_{k},
\end{equation}
the summation is over all $N_{\rm p}$ types of particles, and $\sigma_{j}={\rm sign}(q_{j})$ depend
on $n_{j}$ and $v_{j}$, see\cite{TS72}.

\subsection{Dressing of excitations}
Besides the particle content, the hydrodynamic approach requires to extract energies and momenta of individual particles.
These are dressed by the interaction with a non-trivial vacuum (a reference macrostate).
The dressed energies $\omega_{j}(u)$ and momenta $p_{j}(u)$ of excitations on top of a given macrostate are determined from
\begin{align}
\omega_{j} &= e_{j} + F_{kj}\star \sigma_{k}e_{k}\vartheta_{k},\\
p_{j} &= \theta_{j} + F_{kj}\star \sigma_{k}a_{k}\vartheta_{k},
\end{align}
where $e_{j} \simeq a_{j}$ and $\theta_{j}=\ii \log S_{j}$ are the bare single particle energy and momenta, respectively, whereas
$\vartheta_{k}=\rho_{k}/(\rho_{k}+\bar{\rho}_{k})$ are the filling functions pertaining to the reference macrostate.
Shift functions $F_{kj}(u,t)$ encode the $\mathcal{O}(1/L)$ shift of a rapidity $u$ for a particle of type $k$ caused by
the injection of a particle of type $j$ carrying rapidity $t$,
\begin{equation}
u \to u - \frac{1}{L}\frac{\sigma_{k}F_{kj}(u,t)}{\rho_{k}(u)+\bar{\rho}_{k}(u)},
\end{equation}
and obeys the following integral equation
\begin{equation}
F_{jm}(u,v) = \frac{1}{2\pi \ii}\log S_{jm}(u-v) -
\sum_{k=1}^{N_{\rm p}}\int_{\mathbb{R}} \dd t\,\sigma_{k}a_{jk}(u-t)\vartheta_{k}(t)F_{km}(t,v).
\end{equation}

\subsection{Particle density operators}
Every particle in the spectrum is assigned a particle density operator $\hat{\rho}_{j}(u)$, representing (by definition)
conserved operators whose action on thermodynamic eigenstates return Bethe root distributions $\rho_{j}(u)$. Particle density
operators can thus be perceived as interacting counterparts of the (momentum) mode occupation numbers
in non-interacting theories~\cite{IQC17}.

In the $|\Delta|\geq 1$ regime we put $\Delta = \cosh{(\eta)}=\tfrac{1}{2}(\qdef + \qdef^{-1})$.
The complete set of $\hat{\rho}_{j}(u)$ ($j\in \mathbb{N}$, $u\in \mathbb{R}$) is constructed
from commuting fused transfer operators $\hat{T}_{j}(u)$, $[\hat{T}_{j}(u),\hat{T}_{j^{\prime}}(u^{\prime})]=0$, constructed
in the standard way,
\begin{equation}
\hat{T}_{j}(u)={\rm Tr}_{\mathcal{V}_{j}}\hat{L}^{(1)}_{j}(u)\hat{L}^{(2)}_{j}(u)\cdots \hat{L}^{(L)}_{j}(u),
\end{equation}
where the Lax operators read
\begin{equation}
\hat{L}_{j}(u) = \frac{1}{\sinh{(\eta)}}
\begin{pmatrix}
\sin{(u+\ii\eta \hat{S}^{z}_{j})} & \ii \sinh{(\eta)}\hat{S}^{-}_{j}\\
\ii \sinh{(\eta)}\hat{S}^{+}_{j} & \sin{(u-\ii \eta \hat{S}^{z}_{j})}
\end{pmatrix},
\end{equation}
with the $\qdef$-deformed spin-$j/2$ generators $\hat{S}^{\pm}_{j},\hat{S}^{z}_{j}$ fulfilling the deformed algebraic
relations (writing $\hat{K}_{j}=\qdef^{\hat{S}^{z}_{j}}$),
\begin{equation}
\hat{K}_{j}\hat{S}^{\pm}_{j} = \qdef^{\pm 1}\hat{S}^{\pm}_{j}\hat{K}_{j},\quad
[\hat{S}^{+}_{j},\hat{S}^{-}_{j}] = \frac{\hat{K}^{2}_{j}-\hat{K}^{-2}_{j}}{\qdef - \qdef^{-1}},
\end{equation}
acting on (higher-spin) \emph{unitary} irreducible $(j+1)$-dimensional $\mathcal{U}_{\qdef}(\mathfrak{sl}_{2})$
modules $\mathcal{V}_{j}$ as
\begin{align}
\hat{S}^{z}_{j} &= (j/2-m)\ket{m}\bra{m},\\
\hat{S}^{+}_{j} &= \sqrt{[j-m]_{\qdef}[m+1]_{\qdef}}\ket{m+1}\bra{m},\\
\hat{S}^{-} &= \sqrt{[j-m]_{\qdef}[m+1]_{\qdef}}\ket{m}\bra{m+1},
\end{align}
for $m=0,1,\ldots,j$, and where the $\qdef$-numbers $[x]_{\qdef}=(\qdef^{x}-\qdef^{-x})/(\qdef-\qdef^{-1})$ have been introduced.

Introducing a set of extensive (local) conserved operators (see \cite{IMP15,IlievskiGGE15,StringCharge,IMPZ16_review}),
\begin{equation}
\hat{X}_{j}(u) = \frac{1}{2\pi \ii}\partial_{u}\log \frac{\hat{T}_{j}(u+\tfrac{\ii \eta}{2})}{T_{0}(u+j\tfrac{\ii \eta}{2})},
\end{equation}
defined for spectral parameter $u$ within the `physical strip' $u\in \mathcal{P}$,
\begin{equation}
\mathcal{P} = \left\{u\in \mathbb{C};|{\rm Im}(u)|<\tfrac{\eta}{2}\right\},
\end{equation}
the particle density operators $\hat{\rho}_{j}(u)$ ($j\in \mathbb{N}$) are given by~\cite{StringCharge}
\begin{equation}
\hat{\rho}_{j} = \hat{X}^{+}_{j} + \hat{X}^{-}_{j} - \hat{X}_{j-1} - \hat{X}_{j+1} \equiv \square \hat{X}_{j},
\end{equation}
where $\hat{X}_{0}\equiv 0$ and we have used the compact notation for imaginary shifts,
$f^{\pm}(u)\equiv f(u\pm \tfrac{\ii \eta}{2}\mp \ii 0)$. Notice that (auxiliary) higher-spin irreducible
representations of $\mathcal{U}_{\qdef}(\mathfrak{sl}_{2})$ ($\qdef \in \mathbb{R}$) are in one-to-one correspondence with the 
particle types. Moreover, by virtue of unitarity or $\mathcal{V}_{j}$, the particle operators $\hat{\rho}_{j}(u)$ commute with the 
spin-reversal transformation $\hat{R}=\prod_{i}\hat{S}^{x}_{i}$, $[\hat{R},\hat{\rho}_{j}(u)]=0$.

\subsubsection{Interval $|\Delta|<1$.}
The critical interval is parametrized by $\Delta=\tfrac{1}{2}(\qdef+\qdef^{-1})=\cos{(\gamma)}$. For $\gamma/\pi=m/\ell$ being
a root of unity, the particle spectrum truncates to a finite set. Writing the (truncated) continued fraction expansion,
\begin{equation}
\frac{\gamma}{\pi} = \frac{1}{\nu_{1}+\frac{1}{\nu_{2}+\frac{1}{\nu_{3}+\ldots}}}\equiv (\nu_{1},\nu_{2},\ldots,\nu_{l}),
\end{equation}
the total number of distinct particle types is $N_{\rm p}=\sum_{i=1}^{l}\nu_{i}$.
The complete classification can be found in \cite{TS72}, see also \cite{StringCharge}. In contrast to the $|\Delta|\geq 1$ case,
the number of linearly independent unitary transfer operators $\hat{T}_{j}(u)$ is now finite,
with $j=1,2,\ldots,N_{\rm p}-1$. Moreover, labels $j$ do no longer directly correspond to the sizes of
auxiliary spins, but are instead non-trivially related to the string lengths and parities (cf. \cite{StringCharge}).
The mapping between the particle density operators $\hat{\rho}_{j}(u)$ and a family of (extensive)
conserved operators $\hat{X}_{j}(u)$ for arbitrary $\gamma$ has been derived in \cite{StringCharge}.
Denoting $\hat{\rho}_{N_{\rm p}-1}\equiv \hat{\rho}_{\circ}$ and $\hat{\rho}_{N_{\rm p}}\equiv \hat{\rho}_{\bullet}$, the density 
operators can be given in a covariant form
\begin{align}
\hat{\rho}_{j} &= \square_{\gamma} \hat{X}_{j},\qquad j=1,2,\ldots N_{\rm p}-2,\\
\hat{\rho}_{\circ} - \hat{\rho}_{\bullet} &= \square_{\gamma} \hat{X}_{\ell-1},
\end{align}
with $X_{N_{\rm p}}\equiv 0$, and where $\square_{\gamma}$ is a $\gamma$-dependent discrete wave operator modified introduced
in~\cite{StringCharge}. It is crucial to stress that the spectra of $\hat{X}_{j}(u)$, for $j=1,2,\ldots,N_{\rm p}-1$ only allow 
to determine the densities for $\rho_{j}(u)$ with $j=1,2,\ldots N_{\rm p}-2$, and the \emph{difference} of the `boundary particles'
$\rho_{\circ} - \rho_{\bullet}$. Therefore, in distinction to $|\Delta|\geq 1$, the particle content in the interval $|\Delta|<1$
(at $\qdef$ root of unity) is no longer in  bijection with unitary (i.e. spin-reversal invariant) irreducible representations of the corresponding quantum symmetry $\mathcal{U}_{\qdef}(\mathfrak{sl}_{2})$.
In order retrieve the missing information and obtain $\hat{\rho}_{\circ}$ and $\hat{\rho}_{\bullet}$ 
\emph{separately}, the set $\hat{X}_{j}(u)$ has to be supplemented with an extra conserved
operator $\hat{Z}(u)$ (cf. Eq.~\eqref{eqn:Z-operator} below) built from \emph{non-unitary} auxiliary irreducible representations of 
$\mathcal{U}_{\qdef}(\mathfrak{sl}_{2})$ (with $\qdef$ being a root of unity), constructed first in \cite{Prosen11,PI13} (see
also~\cite{Prosen14,Pereira14}). The distinguished property of $\hat{Z}(u)$ is that it flips the sign under spin-reversal 
transformation, $\hat{R}\hat{Z}(u)\hat{R} = -\hat{Z}(u)$.

For instance, in the simplest case of $\gamma/\pi = 1/\nu_{1}=1/\ell$, the lengths and parties of particles are
\begin{align}
n_{j} &= j,\quad v_{j}=1,\qquad j=1,2,\ldots,\nu_{1}-1,\\
n_{\nu_{1}} &= 1,\quad v_{\nu_{1}} = -1.
\end{align}
Here $\square_{\gamma}$ coincides with the gapped counterpart $\square$, with the imaginary shifts given now by $\ii \gamma/2$.
At the boundary nodes we have~\cite{DeLuca16}
\begin{equation}
\hat{\rho}_{\bullet} = -\frac{1}{2}\left(\hat{X}^{+}_{\ell-1} + \hat{X}^{-}_{\ell-1}\right)
-\frac{1}{2\gamma}\int_{-\gamma/2}^{\gamma/2}\dd z\,\hat{X}^{\prime}(u+\ii z),
\end{equation}
where
\begin{equation}
\hat{X}^{\prime}(u) = \partial_{\alpha}\hat{X}_{\ell-1,\alpha}(u)|_{\alpha=0} = \hat{Z}(u) - \frac{\gamma}{2\pi \cosh^{2}{(u)}}.
\label{eqn:Z-operator}
\end{equation}
Here $\hat{X}_{\ell-1,\alpha}(u)$ is defined as a conserved operator built from the (finite-dimensional) non-unitary irreducible
representation $\mathcal{V}_{\ell-1,\alpha}$, with the action of $\qdef$-deformed spin operators reading~\cite{DeLuca16}
\begin{align}
\hat{K}_{\ell-1,\alpha}\ket{m} &= \qdef^{m+\alpha}\ket{m},\\
\hat{S}^{+}_{\ell-1,\alpha}\ket{m} &= -[m-\ell+1+2\alpha]_{\qdef}\ket{m+1},\\
\hat{S}^{-}_{\ell-1,\alpha}\ket{m} &= [m+\ell-1]_{\qdef}\ket{m-1},
\end{align}
where $m=-\tfrac{1}{2}(\ell-1),\ldots,\tfrac{1}{2}(\ell-1)$.
The upshot of this is that macrostates for which $\rho_{\circ}(u)\neq \bar{\rho}_{\bullet}(u)$ carry non-vanishing amount of
$Z$-charge, implying that the spin reversal operation yields a (locally) distinguishable macrostate (i.e. a state
with distinct particle densities). In the context of our application, this enables a ballistic drift of particles across a
magnetic domain wall.

\clearpage
\section{Additional plots}

\begin{figure}[h]
\includegraphics[width=0.6\textwidth]{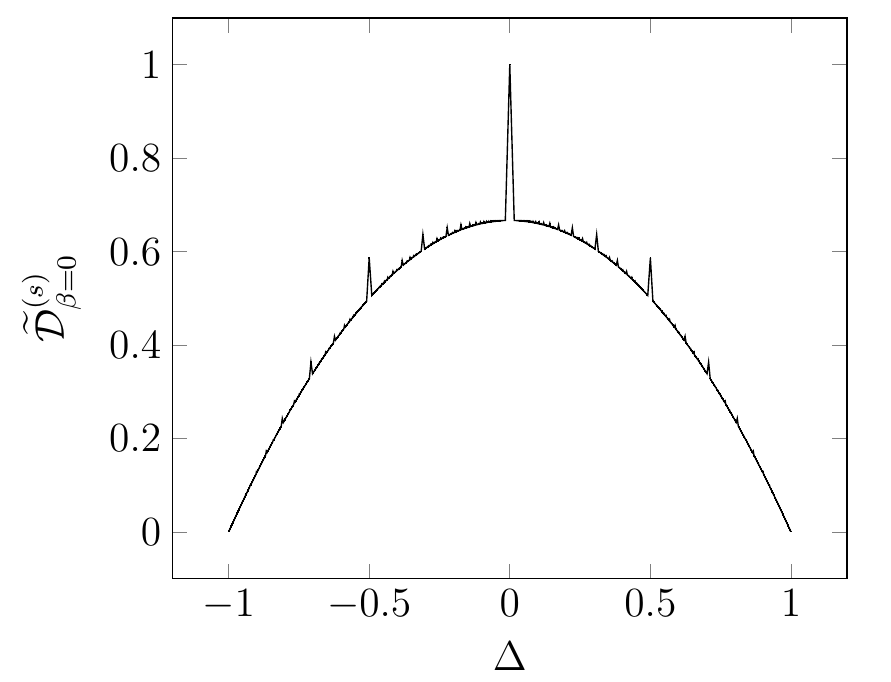}
\caption{Rescaled spin Drude weight $\widetilde{\DD}^{(s)} = (16/\beta)\DD^{(s)}$ in the high-temperature $\beta \to 0$ limit
as function of anisotropy $\Delta$, given by analytic expression, equation (5) in the main text.
The result was obtained in \cite{PI13} as a lower bound for the spin Drude weight.
In the Letter we rigorously prove that the bound is optimal and coincides with the exact value of $\DD^{(s)}$ at
infinite temperature.}
\end{figure}

\begin{figure}[h]
\includegraphics[width=0.6\textwidth]{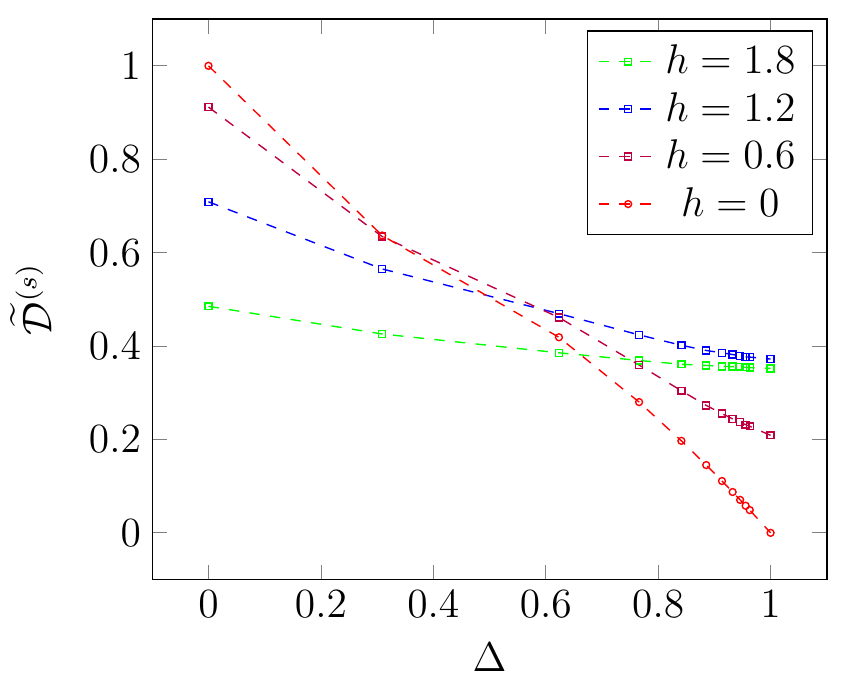}
\caption{Rescaled spin Drude weight $\widetilde{\DD}^{(s)} = (16/\beta)\DD^{(s)}$ at $\beta \to 0$ as function of
anisotropy $\Delta$ for $\Delta=0$ and $\Delta=\cos{(\frac{\pi}{\nu_{1} + 1/2})}$ with $\nu_{1} = \{2,3,\ldots,12\}$ and
$\nu_{2} \to \infty$ (obtained from the previous $\nu_{2}$-points by linear fitting), shown for various chemical potentials $\mu_s=h$.}
\end{figure}

\begin{figure}[h]
\includegraphics[width=0.6\textwidth]{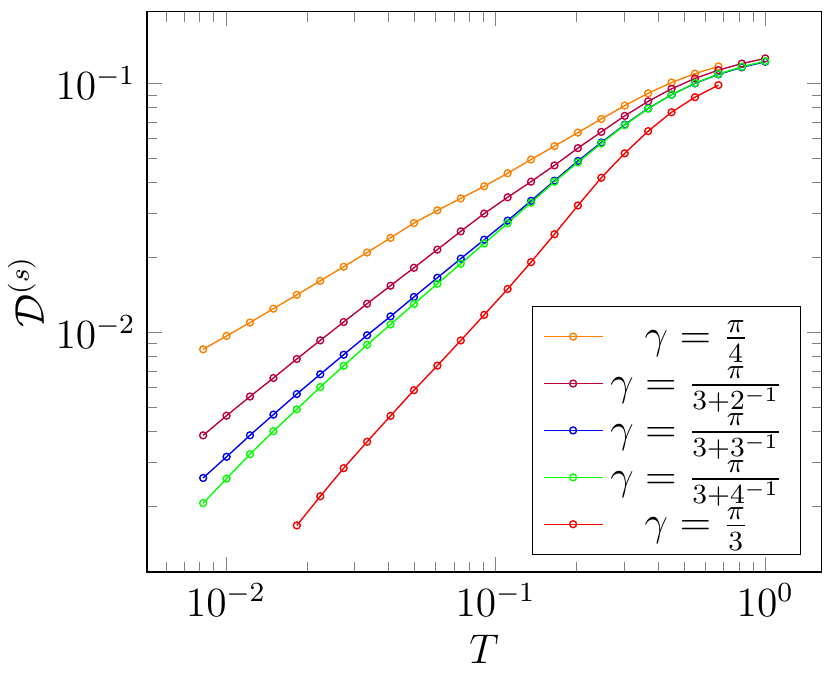}
\caption{Log-log scale: Spin Drude weight $\DD^{(s)} -  \DD^{(s)}(T=0)$ as function of temperature $T=\beta^{-1}$ at 
$\Delta=\cos{(\gamma)}$. For $\gamma=\pi/\nu_{1}$ we confirm the scaling
$\DD^{(s)} -  \DD^{(s)}(T=0)\sim T^{2/(\nu_{1}-1)}$ found earlier in~\cite{Zotos99}, while at $\gamma/\pi = 1/(\nu_{1} + 1/(\nu_{2} + 1/\nu_3))$
we observe the power law $\DD^{(s)} -  \DD^{(s)}(T=0)\sim T^{2/(\gamma/\pi -1)}$.}
\end{figure}

\begin{figure}[h]
 \includegraphics[width=0.6\textwidth]{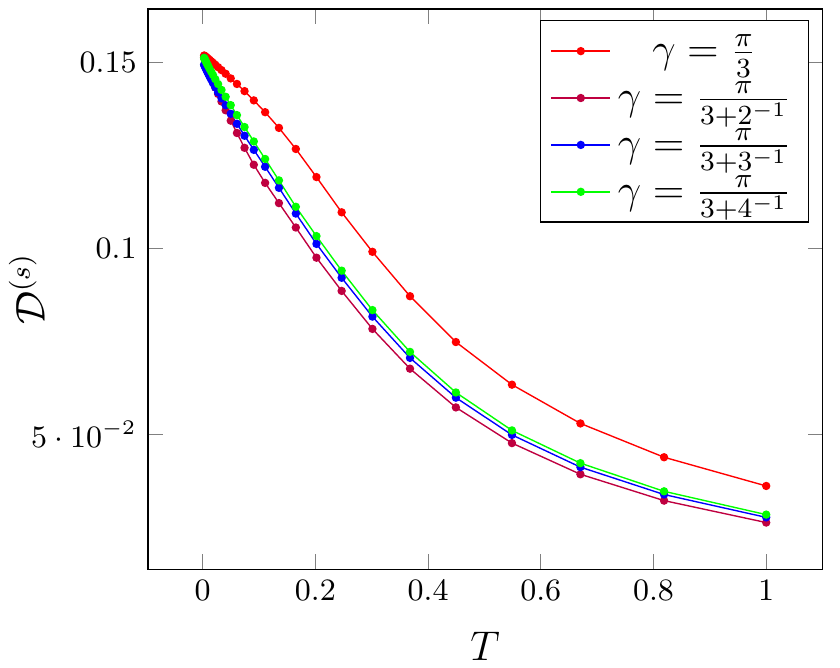}
\caption{Spin Drude weight  $\DD^{(s)}$ as function of temperature $T=\beta^{-1}$, shown for
$\Delta=\cos{(\gamma)}$ at $\gamma/\pi=1/(3 + 1/\nu_2)$ for $\nu_2 = \{2,3,4,\ldots\}$.
While in the low-$T$ regime $\DD^{(s)}$ for $\Delta=\cos{(\pi/(3 + 4^{-1}))} \approx 0.568$ and $\Delta=0.5$ are
comparable to each other, they significantly differ at higher temperatures.}
\end{figure}

\end{document}